\documentclass[twocolumn, aps, pra, superscriptaddress]{revtex4-1}
\pdfoutput=1
\usepackage[utf8]{inputenc}
\usepackage[english]{babel}
\usepackage[T1]{fontenc}
\usepackage{hyperref}
\usepackage{graphicx}
\usepackage{amsfonts, amsmath, amsthm, amssymb} 
\usepackage{braket}
\usepackage{float}
\usepackage{amsthm}
\usepackage{color}
\usepackage[all]{nowidow}
\usepackage[ruled,lined,linesnumbered]{algorithm2e}
\usepackage{xurl}
\usepackage{changes}

\begin{document}

\title{Bounding Large-Scale Bell Inequalities}

\author{Luke Mortimer}
\affiliation{ICFO-Institut de Ciencies Fotoniques, The Barcelona Institute of Science and Technology, 08860 Castelldefels, Spain}

\date{\today}

\begin{abstract}
\noindent
Bell inequalities are an important tool for studying non-locality, however quickly become computationally intractable as the system size grows. We consider a novel method for finding an upper bound for the quantum violation of such inequalities by combining the NPA hierarchy, the method of alternating projections, and the memory-efficient optimisation algorithm L-BFGS. Whilst our method may not give the tightest upper bound possible, it often does so  several orders of magnitude faster than state-of-the-art solvers, with minimal memory usage, thus allowing solutions to problems that would otherwise be intractable. We benchmark using the well-studied $I_{3322}$ inequality as well as a more general large-scale randomized inequality $R_{xx22}$. For randomized inequalities with 130 inputs either side (a first-level moment matrix of size 261x261), our method is $\sim 100$x faster than both MOSEK and SCS whilst giving a bound only $\sim 2\%$ above the optimum.
\end{abstract}

\maketitle


\section{Introduction}

In 1964, John Bell first introduced his response to the so-called Einstein-Podolsky-Rosen paradox, showing that one could devise an experiment in which one, based only on measuring a system, could prove that the resulting output statistics do not belong to a model with only local hidden variables \cite{Bell1964}. The now well-established format of such experiments is to show that some statistics violate a Bell inequality: a linear combination of expectation values that in the purely classical case are bounded by some limit, but violated in the case that the system is quantum and displays non-locality.

Such inequalities play a crucial role in many quantum information protocols as they are capable of certifying the ``quantum-ness'' of a system. They have uses in quantum key distribution \cite{acin2006efficient,masanes2011secure}, quantum computing \cite{morikoshi2006information}, quantum randomness \cite{dhara2013maximal,fehr2013security} and even more fundamental questions like mutually unbiased bases \cite{tavakoli2021mutually}. However, as the size of the system increases (i.e. more parties, more measurements, more outcomes) the difficulty of finding a set of valid quantum expectation values that maximally violate the inequality quickly becomes intractable. For a review on non-locality, see \cite{brunner2014bell}.

The simplest example is the CHSH inequality, where two parties Alice and Bob share a quantum state $\rho$, each with measurement settings $x={1,2}$ and $y={1,2}$, and each with possible outcomes $a={-1,1}$ and $b={-1,1}$, respectively. It can be thought of as a game, such that each iteration Alice chooses (by some strategy) $x$ to be either $1$ or $2$ corresponding to measuring with measurement $A_1$ or $A_2$, with the binary result of said measurement, $a$, being either $-1$ or $1$, similarly for Bob. Assuming one then measures $a$ and $b$ given a choice of $x$ and $y$, the probability is recorded as $p(ab|xy)$. The expectation values for a given set of measurements therefore takes the form:

\begin{equation}
\begin{aligned}
    \braket{A_xB_y} = \sum_{a=\pm1,b=\pm1} a ~ b ~ p(ab|xy)
\end{aligned}
\end{equation}

In the notation of expectation values, the CHSH inequality is given as:
\begin{equation}
    \braket{A_1 B_1} + \braket{A_1 B_2} +  \braket{A_2 B_1} - \braket{A_2 B_2} \le  2
\end{equation}

such that with purely classical states one can reach a maximum value of 2, whilst with a shared quantum state one can violate the inequality, reaching a maximum quantum value of $2\sqrt{2}$, known as the Tsirelson bound \cite{cirel1980quantum}.

\begin{figure}
    \centering
    \includegraphics[width=0.95\linewidth]{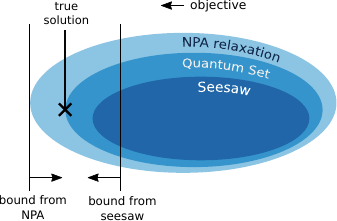}
    \caption{Diagram showing how the seesaw method and the NPA hierarchy create lower and upper bounds on the maximum violation of a Bell inequality. Note that since the seesaw is generally non-convex, it is unlikely that its optimum will be found. As one increases the the level of the NPA hierarchy or the dimension/optimality of the seesaw, each bound should tend towards the true solution.}
    \label{fig:seesaw}
\end{figure}

Perhaps the two most common methods for finding bounds on the quantum value of a given Bell functional are the seesaw method \cite{colomer2022three} and the NPA hierarchy \cite{navascues2008convergent}. In the seesaw, one considers a state of some fixed dimension and then proceeds to optimize the bilinear objective, finding a valid set of quantum measurements that provide a lower bound for the maximum violation. With the NPA hierarchy one relaxes the problem, linearizing the objective and considering a super-set of all possible expectation values over all states of unconstrained dimension, then forming a hierarchy of semidefinite programs (SDPs) that constrain said expectation values to become more and more physical. A diagram displaying the difference between the two methods is given as Figure \ref{fig:seesaw}. In this work we consider the latter case, since it provides an upper bound which is generally the more relevant quantity.

\section{Semidefinite Programming}

SDPs are a highly versatile tool in the world of optimisation, capable of representing all problems in the NP class \cite{o2008optimal} and in many cases providing an efficient method of solving such problems rather than linear programming (a restricted case of semidefinite programming). Whilst here we refer to SDPs with the application of finding bounds for Bell inequalities, they also have applications in portfolio optimization \cite{leibfritz2009successive}, structural engineering \cite{ohsaki1999semi}, machine learning \cite{de2007deploying} and robotics \cite{wu2021sdp}.

The general form of an SDP is to maximize  a linear objective function with the constraints that the matrix of variables obeys a series of linear relations as well as being positive (having all eigenvalues non-negative):
\begin{equation}
    \begin{aligned}
        \max_{X} \quad &\braket{C, X}_F \\
        \text{s.t.} \quad &\braket{A_i, X}_F = b_i \quad \forall i=1,\ldots,m \\
        &X \succeq 0
    \end{aligned}
    \label{eq:sdp}
\end{equation}

A number of methods exist to solve SDPs, perhaps the most popular being the idea of an interior point solver. This type of method first finds a point inside the feasible region (satisfying the constraints) and then tries to follow an optimal path through the space to reach the solution. How it calculates this optimal path varies between implementations: first-order solvers simply following the gradient of the Lagrangian, resulting in slower (linear) convergence but being relatively faster per iteration (at least linear in the number of variables), whilst second-order solvers calculate the Hessian (matrix of second-order derivatives) to reach much faster (quadratic) convergence at the cost of a much more intensive per iteration calculation (at least quadratic in the number of variables). Second-order solvers also use large amounts of memory to store the Hessian. Example of first-order interior point solvers include SCS \cite{o2016conic} and SDPA \cite{yamashita2003implementation}, whilst a common second-order solver is MOSEK \cite{andersen2000mosek} which is often the fastest to converge for many problems of interest \cite{bench}.

For any optimisation problem there exists the concept of duality \cite{boyd2004convex}, such that the dual of a primal problem approaches the solution from ``the other side''. In the case of the NPA hierarchy, a convex maximization problem giving an upper bound on a quantity, the dual is a convex minimization problem such that each point inside its feasible set is a valid upper bound for the quantity, with the optimum of the dual being the lowest possible upper bound. As such, one can argue that the dual of the NPA is perhaps the more useful formation, since any interior point provides a valid bound, whilst a random interior point of the primal provides no useful information (it is a lower bound to the upper bound, but not necessarily a lower bound to the true solution). The dual of an SDP in the form of Equation \ref{eq:sdp} is given by the following, where $y$ is a new vector containing a variable for each of the original linear constraints:
\begin{equation}
    \begin{aligned}
        \min_{y} \quad &b \cdot y \\
        \text{s.t.} \quad &C - \sum_i^m A_i y_i \succeq 0 
    \end{aligned}
    \label{eq:dual}
\end{equation}


\section{Our Method}

Consider the case where we have an SDP corresponding to a certain level of the NPA hierarchy. This SDP has an optimum which provides an upper bound for the quantum bound (sometimes referred to as ``the value'', for simplicity) of the Bell inequality. We first begin by taking the dual of this SDP, resulting in a new SDP whose optimum also provides an upper bound for the value, except that so too does every point in the feasible set (albeit worse bounds than the optimum). We therefore change to considering the relaxed problem of finding a point inside the feasible set as fast as possible, ideally as close to the optimum as possible.

\begin{figure}[h]
    \centering
    \includegraphics[width=0.95\linewidth]{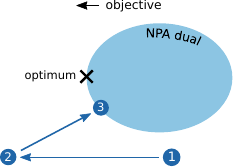}
    \caption{Diagram showing the outline of our method. Beginning with the dual problem of a given level of the NPA hierarchy, one starts at a point (point 1), generally from zero, then travel very far in the direction of the objective (point 2), before projecting back onto the set (point 3), giving an interior point near the objective and providing an upper bound for the Bell inequality.}
    \label{fig:exile}
\end{figure}

The core of our idea is as follows: one travels as far away from the feasible set as possible in the direction of the objective function, ideally to the limit of infinity but in our case always truncating to some large (relative to the problem) distance. Then one projects back onto the set using the method of alternating projections, accelerated using the L-BFGS algorithm. This rapidly gives a feasible point on the edge of the set, ideally the nearest point to the far-away point (the optimum) but often simply a point nearby. A diagram showing this idea of ``exile and projection'' is given as Figure \ref{fig:exile}

Although this method might not give the best upper bound possible, it does so in a fraction of the time and memory than solving the SDP (primal or dual) with a state-of-the-art SDP solver. If one desires a better bound, one can then travel outside the set towards the objective function and repeat the projection, eventually converging (albeit slowly) to the optimum, however in general the focus of this work is to get a quick-and-dirty bound as fast as possible, rather than strictly the optimum.

\subsection{Method of Alternating Projections}

We now consider the problem of trying to find the closest point in a set $A$ to a given point $y$ (``projecting $y$ onto $A$''). For many simple convex domains this can be done analytically, normally by first considering the problem:
\begin{equation}
    \begin{aligned}
        \min_x \quad &||x-y||^2 \\
        s.t. \quad &x \in A
    \end{aligned}
\end{equation}

One then forms the Lagrangian and solves for the KKT conditions \cite{gordon2012karush}. In the case of the affine set $Ax=b$, the projection $P$ is given by the pseudo-inverse \cite{boyd2004convex}:
\begin{equation}
    P_{\text{affine}}(y) = y - A^T(AA^T)^{-1}(Ay-b)
\end{equation}

In the case of projecting a matrix onto the positive semidefinite (PSD) cone, the projection can be found by simply taking the eigenvalues of the original matrix, setting all negative eigenvalues to zero, then reconstructing the matrix \cite{boyd2004convex}, such that if the eigenvalues/vectors of $y$ are $\lambda_i$ and $\ket{\phi_i}$, one has 
\begin{equation}
    P_{\text{PSD}}(y) = \sum_i \text{max}(\lambda_i,0) \ket{\phi_i}\bra{\phi_i}
\end{equation}

Whilst projecting onto textbook-friendly sets like these is easy, projecting onto their intersection can be considerable harder (although still ``easy'' in a complexity sense, since SDPs are in P even if they often represent relaxations of problems seemingly not in P). The simplest method of doing this is the method of alternating projections, whereby one first projects onto one set, then the other, then repeat until one has a point in the intersection. This method is widely used in quantum information theory \cite{barbera2025boosting,ticozzi2017alternating,drusvyatskiy2015projection}, and as such any improvement that can be found is of general interest.

\begin{figure}[h]
    \centering
    \includegraphics[width=0.95\linewidth]{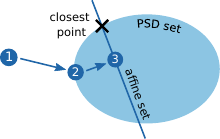}
    \caption{Diagram showing how the method of alternating projections might not always find the closest point in the set to the point being projected (point 1). By projecting first onto the PSD set (point 2) we move away from the true projection such that when we then project onto the affine set (point 3) we do not get the closest point. Whilst in this diagram it appears that one could simply project onto the affine set first to get the optimum, that is generally not the case.}
    \label{fig:alternating}
\end{figure}

Convergence to a point in the set is guaranteed if the sets are convex and their intersection is non-empty, although it won't necessarily be the closest point. An example of how this can occur is shown in Figure \ref{fig:alternating}. Whilst there does exist a modification of the alternating projection method which promises to always converge to the closest point, known as Dykstra's projection algorithm \cite{bauschke1994dykstra}, when the point is very far from the set this seems to always take a long time to converge and thus for our purposes offers very poor performance. 

\subsection{L-BFGS}

The alternating projection algorithm is known to, in general, converge linearly. Thus the error is proportional to the previous error, e.g. $\epsilon_i = 0.7 \epsilon_{i-1}$. This is okay for many applications, however in our case it often required several hundred iterations to reach a tolerance of $10^{-8}$. However, upon doing principal component analysis (PCA) on the sequence produced we noticed that the points follow a very smooth path, and thus may be susceptible to some kind of convergence acceleration. We first tested using Aitken's delta-squared process \cite{overholt1965extended}, which provided a minor speed-up (e.g. from 300 iterations to 250), but we then found much greater success applying an algorithm allowing quadratic convergence.

In order to reach quadratic convergence (such that the error is proportional to the square of the previous error, e.g. $\epsilon_i = 0.7 \epsilon_{i-1}^2$), one possible method is to use the full Hessian, in the style of Newton's method for root-finding \cite{hansen1976family}. This, however, requires a large number of derivative calculations, which often proves expensive in both time and memory. An important advancement was the use of the approximate Hessian by Broyden, Fletcher, Goldfarb and Shanno (BFGS), allowing a much faster iterate \cite{fletcher2000practical}. This was then further improved in the limited-memory version (L-BFGS) by instead finding an approximate Hessian using only a few (often less than 10) previous gradients, whilst still maintaining super-linear convergence  \cite{byrd1995limited}.

\begin{figure}[h]
    \centering
    \includegraphics[width=0.95\linewidth]{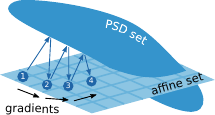}
    \caption{Diagram in 3D showing how we get the gradients for use with the L-BFGS algorithm. By projecting first onto the PSD set and then the affine set, we consider the difference between affine points as our gradients, which we then use to estimate the Hessian and thus accelerate convergence. As an aside, several proof-readers noted that this diagram may somewhat resemble a whale when viewed from afar.}
    \label{fig:lbfgs}
\end{figure}

One might note however, that we don't have an explicit nor smooth form for the gradient of our optimisation, since we are simply alternating projections. The method in which we approximate the gradient and thus a Hessian is by considering only the iterations in which we project onto the affine set. That is, we begin at a point in the affine set, project onto the semidefinite cone and then back to the affine set. The ``gradient'' is then simply the difference between the two points. A diagram is given as Figure \ref{fig:lbfgs}. By feeding this gradient vector into an existing implementation of L-BFGS as well as using the squared error as the ``objective'' for the L-BFGS line search, we get convergence in super-linear time at minimal memory usage. What previous required hundreds of iterations now requires only a few, at almost no extra computational cost.

\subsection{Improving the bound}

Given the ability to find a projection at a point in the intersection very quickly as a black-box, it is possible to then get a point closer to the original point outside the set. The best method we found to do this is to simply travel from the projected point some fraction of the distance towards the point outside the set. For instance, assuming we start at a point $y$ and then have iterates $x_i$, it should be relatively clear that if we move an infinitesimal distance towards $y$ and then project back onto the set, we get a resulting point also on the edge of the set that will be slightly closer to $y$. For faster convergence we take larger steps, reducing geometrically at a ``slow enough'' rate: $x_{i+1} = P(x_i + \alpha (y - x_i))$ such that each $\alpha_{i+1} = c \alpha_i$. Here $c$ is some coefficient chosen to determine the rate of decrease of the step size, perhaps similar in spirit to methods such as simulated annealing which converge to the optimum if the system is ``cooled'' slow enough.

We have tested variations on this improvement procedure significantly, including using such steps to get an ``outer'' gradient and then feeding that into L-BFGS, using Aitken's delta-squared process, as well as testing various gradient descent algorithms, however we found no significant improvement. We also tested adding an extra affine constraint constraining that the objective equals a certain (lower) value, but found that this prevents L-BFGS from working efficiently and thus it seems that the super-linear convergence is restricted to just the standard moment-matrix set of constraints. 

\subsection{Implementation}

Our implementation of this algorithm, as well as the test suite benchmarking against other algorithms, is all written in C++. For the implementation of L-BFGS we use Optim \cite{mogensen2018optim}. For all matrix/vector operations we use the C++ library Eigen \cite{eigenweb}. In order to find the projection onto the affine set, whilst one could calculate and apply the pseudo-inverse, in our tests this often takes longer than using Eigen's implementation of BiCGSTAB \cite{sleijpen1994bicgstab} to attempt to solve the set of linear constraints whilst using the point to be projected as the warm start. To find the projection onto the PSD set, we use Eigen's Self Adjoint Eigensolver to get the eigenvectors and eigenvalues, which we then use as described earlier to get the projection.

To further summarize the idea of the method, we provide some pseudocode, given as Algorithm \ref{alg:summary}. The code we use to generate all data in this work is open source and available on GitHub \cite{code}.

\begin{algorithm}
\SetInd{0.2em}{1.3em}
\DontPrintSemicolon
\SetAlgoLined
\caption{Summary}
\label{alg:summary}
prepare the linear solver \;
initialize the distance \;
travel in the direction of the objective \;
project onto the affine space \;
\Repeat{\textup{enough iterations performed}}{
    project onto the positive cone \;
    project onto the affine space \;
    update the approx Hessian using this new point \;
    use the approx Hessian to get the next point \;
}
\If{\textup{better solution desired}} {
    reduce the distance \;
    go to line 4, keeping the current point \;
}
\end{algorithm}

The following benchmarks were all performed with the same desktop computer: An Intel(R) Xeon(R) W-2295 CPU @ 3.00GHz (18 physical cores, 36 threads) with 128GB RAM. The method is fully parallelized due to the two slowest parts of the algorithm, the projections, already having parallelized implementations in Eigen.


\section{Benchmarks}

\subsection{The $I_{3322}$ Inequality}

As a proof-of-principle, we begin with the smallest (in terms of number of inputs) inequality  not reducible to the CHSH inequality, known as the $I_{3322}$ inequality \cite{collins2004relevant,pal2010maximal}. Here both Alice and Bob each can choose between three two-outcome measurements (hence the 3-3-2-2 part of the name), such that there exists a set of measurements acting on a shared quantum state that display non-locality by violation of a classical bound, as with CHSH. A more interesting property of $I_{3322}$ is its ``hardness'' despite only having a very slightly more complex description than CHSH: in order to fully characterize the quantum bound using NPA, at least level 4 of the hierarchy is required (only level 1 is required with CHSH) \cite{navascues2008convergent,pal2010maximal}, and to find a lower bound using seesaw one is required to use a shared state more complex than a qubit \cite{navascues2015characterizing}. In this work we consider the -1/1 formulation of this inequality, given as:
\begin{equation}
    \begin{aligned}
    &\braket{A_1}-\braket{A_2}+\braket{B_1}-\braket{B_2}-\braket{A_1B_1}\\
    &+\braket{A_1B_2 }+\braket{A_2B_1}-\braket{A_2B_2}+\braket{A_1B_3}\\
    &+\braket{A_2B_3}+\braket{A_3B_1}+\braket{A_3B_2} ~\le~ 4
\end{aligned}
\end{equation}

This has a maximum quantum violation of $5.5$ at level 1, $5.00376$ at level 2 and $5.0035$ at level 3. We begin by showing in Table \ref{tbl:i3322-l1} that our method gives a valid upper bound with a single shot with a fraction of the resources of the other solvers, and then proceeds to converge to the optimum within precision (linear error less than $10^{-10}$) using only a further 12 iterations of refinement. As will be demonstrated later, the initial bound being quite loose appears to be a feature notable of smaller problems, where perhaps one needs to have a very specific set of variable values in order to even get close to the solution.

\begin{center}
\begin{table}
\begin{tabular}{ |c|c|c|c|c| } 
    \hline
    Solver & Bound & Time & Memory \\ 
    \hline
    MOSEK & 5.5 & 36 ms & 26 MB \\ 
    SCS & 5.5 & 2 ms & 26 MB \\ 
    Ours (1) & 6.34423 ($+15\%$) & 1 ms & 14 MB \\ 
    Ours (12) & 5.5 & 5 ms & 14 MB \\ 
    \hline
\end{tabular}
\caption{Table showing how our method performs against some common solvers when applied to level 1 of the NPA hierarchy for the $I_{3322}$ inequality. The number given after ``Ours'' is the number of iterations performed, whereby 1 corresponds to just a single shot and higher numbers correspond to repeated iterations to improve the bound.}
\label{tbl:i3322-l1}
\end{table}
\end{center}

The main advantage of our method only begins to show as the size of the matrices grows. When now considering level 3 of the hierarchy, in which now the matrices are 88x88 rather than 7x7 as with level 1, we see that the time to do a single shot does not grow so much despite now giving a bound below that of level 1. Performing further iterations then brings the bound much closer at the expense of spending a similar amount of time as with the other solvers. These results are shown in Table \ref{tbl:i3322-l3}.


\begin{center}
\begin{table}
\begin{tabular}{ |c|c|c|c|c| } 
\hline
Solver & Bound & Time & Memory \\ 
\hline
MOSEK & 5.0035 & 17790 ms & 560 MB \\ 
SCS & 5.0035 & 13070 ms & 45 MB \\ 
Ours (1) & 5.41829 ($+8\%$) & 217 ms & 46 MB \\ 
Ours (100) & 5.03148 ($+0.5\%$) & 12540 ms & 46 MB \\ 
\hline
\end{tabular}
\caption{Table showing how our method performs against some common solvers when applied to level 3 of the NPA hierarchy for the $I_{3322}$ inequality. The number given after ``Ours'' is the number of iterations performed, whereby 1 corresponds to just a single shot and higher numbers correspond to repeated iterations to improve the bound.}
\label{tbl:i3322-l3}
\end{table}
\end{center}

\subsection{The $R_{xx22}$ Inequality}

To demonstrate the scalability of the method, we turn to bigger Bell inequalities. We consider a generalized inequality with $x$ inputs either side, but still with two-outcomes for simplicity of generation, containing all one and two-body expectation values, whose coefficients are then randomized uniformly between -1 and 1. An example of such an inequality for $x=2$ inputs either side might be:
\begin{equation}
\begin{aligned}
&0.23\braket{A_1}+0.14\braket{A_2}-0.9\braket{B_1}\\
&+0.82\braket{B_2}-0.68\braket{A_1B_1}-0.2\braket{A_1B_2}\\
&+0.11\braket{A_2B_1}+0.72\braket{A_2B_2}~\le~ C
\end{aligned}
\end{equation}

where $C$ is now some classical bound that can be found by optimizing over the local set, which can be phrased as a similar problem as in the quantum case except with commuting variables. Finding the value of $C$ is not the objective of this work, but theoretically our method can also be applied in the case of any SDP, commuting or not.

\begin{figure}
    \centering
    \includegraphics[width=0.95\linewidth]{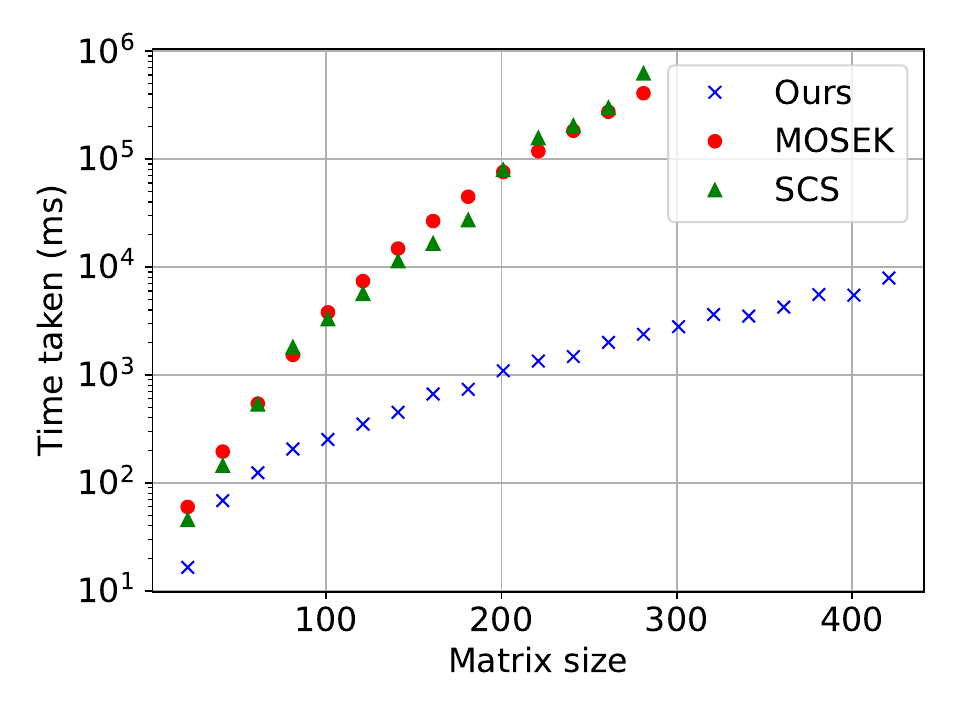}
    \caption{Graph showing how the runtime of our algorithm scales with the size of the moment matrix versus other solvers. Each point is averaged over 100 runs with unique seeds. Error bars not shown because they would be barely visible.}
    \label{fig:time}
\end{figure}

Our main advantage is the time required to find an interior point. As shown by Figure \ref{fig:time}, our method finishes orders of magnitudes faster than the other solvers for larger problems. This is due to the speed of projection combined with the super-linear convergence from L-BFGS.

\begin{figure}
    \centering
    \includegraphics[width=0.95\linewidth]{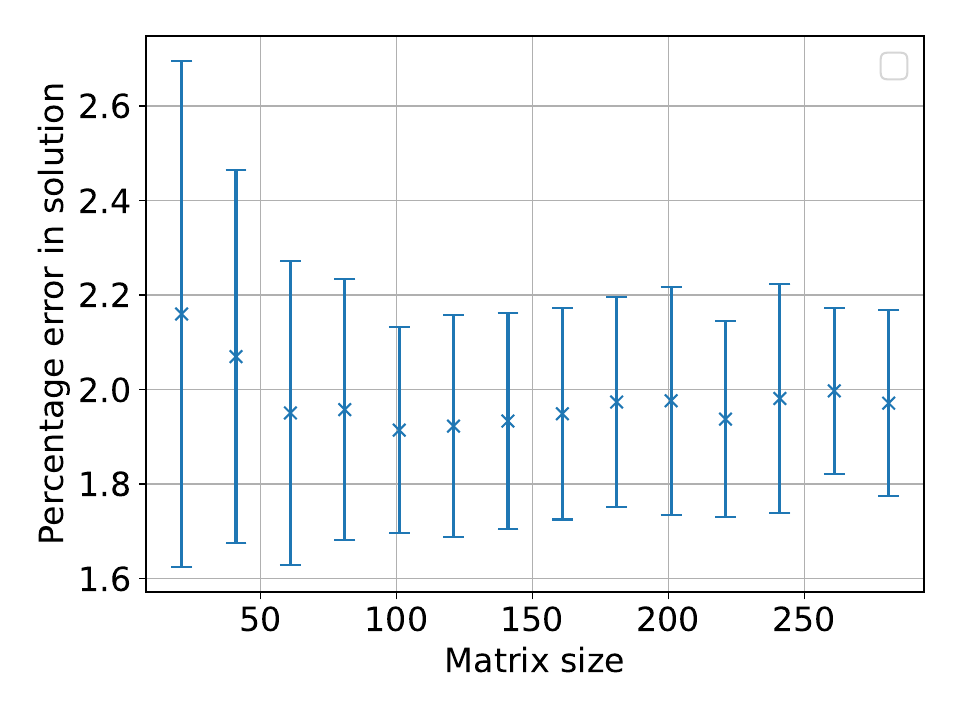}
    \caption{Graph showing how the percentage error of our algorithm scales with the size of the moment matrix. Other solvers can be said to effectively have zero relative error, since they always converge to the optimum (within precision). Each point is averaged over 100 runs with unique seeds, with the standard deviation shown as error bars.}
    \label{fig:error}
\end{figure}

However, such a speed advantage is only useful if the resultant bound is not so far off the optimum. In Figure \ref{fig:error} we show that the percentage error in the solution (the upper bound given by our method versus that of MOSEK or SCS) is generally constant at around $2\%$, seemingly more consistent for larger problems. Thus, if one has a large Bell inequality in which they need a valid bound that needs not be tight, our method demonstrates a clear advantage.

\begin{figure}
    \centering
    \includegraphics[width=0.95\linewidth]{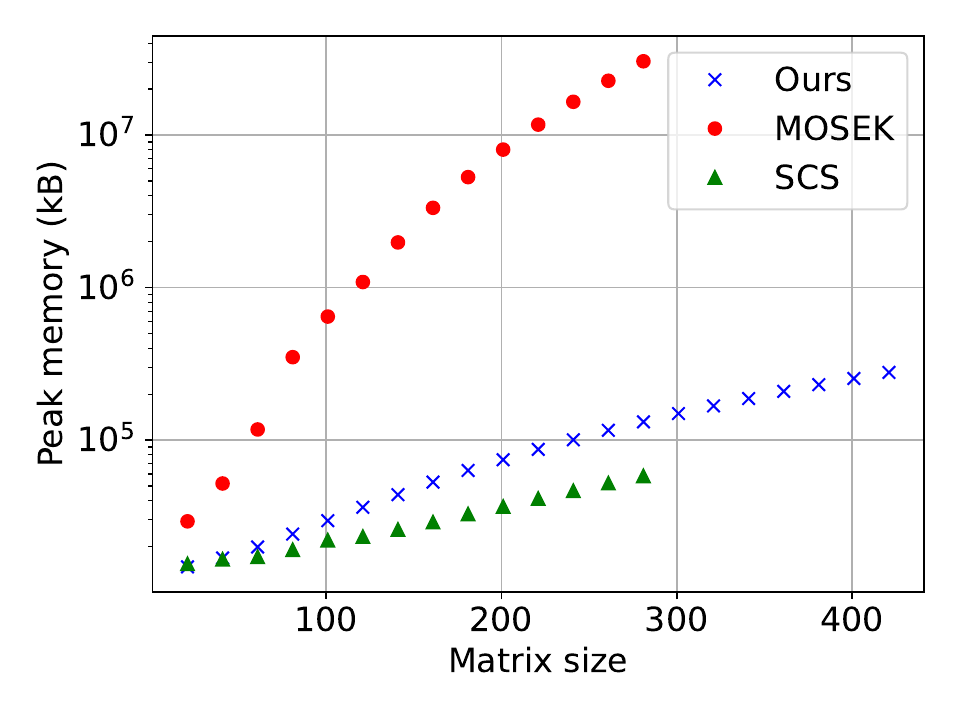}
    \caption{Graph showing how the memory usage of our algorithm scales with the size of the moment matrix versus other solvers. Each point is averaged over 100 runs with unique seeds. Error bars not shown because they would be barely visible.}
    \label{fig:memory}
\end{figure}

We also analyse the peak memory usage of the algorithm, demonstrated in Figure \ref{fig:memory}, in which our method uses memory comparable to that of the other first order solver SCS. Second-order solvers like MOSEK require large amounts of memory to store the large matrices of second derivatives, whereas first-order methods have no such necessity. Our method combines the two, using L-BFGS to only maintain a small fraction of the information required to estimate the second-order gradients.

The main drawback of the method is the looseness of the bounds, which unfortunately prevents the possibility of ever beating true SDP solvers if one wants the tightest bound possible. An extensive search over $R_{xx22}$ inequalities was performed and in no case for the same computational cost (i.e. our method at a higher level and MOSEK at a lower level) did our method find a tighter bound. As such, the main use case appears limited to very large-scale systems in which one cannot even run the level 1 SDP using standard solvers.


\section{Conclusion}

In this work we describe and then demonstrate a novel algorithm for finding upper bounds of the maximum quantum violation of Bell inequalities. The method combines the method of alternating projection with the memory-efficient L-BFGS optimisation algorithm, allowing fast convergence to an interior point. Testing this method with the $I_{3322}$ inequality we show that it gives results comparable with existing solvers. When applied to a large-scale randomized inequality that we dub $R_{xx22}$, it gives bounds generally within $2\%$ error in a fraction of the time and memory of other solvers. Such a method would prove useful in the case in which one needs a quick-and-dirty bound on any Bell inequality that one cannot otherwise optimize due to its size. The technique may also have applications when solving other SDP relaxations or when one wants to project onto large spectahedra.

Future work applications include applying it to specific Bell inequalities known to be hard, applying it to the local set to get the classical bounds, or applying the method to problems other than Bell inequalities. Given that this method is an approximation algorithm for an NP-hard problem, it has a wide range of applications. Further work also needs to be done to improve the ability of successive iterations to converge to the optimum, either with a smarter choice of $\alpha$ scaling or using another method entirely.


\section{Acknowledgements}

Thanks to Leonardo Zambrano, Hippolyte Dourdent and Antonio Acín for the useful discussions.

This  project  has  received  funding  from  the  European  Union’s  Horizon  2020  research and innovation programme under the Marie Skłodowska-Curie grant agreement No 847517, the Government of Spain (Severo Ochoa CEX2019-000910-S and FUNQIP), Fundació Cellex, Fundació Mir-Puig and the Generalitat de Catalunya (CERCA program).

\bibliographystyle{unsrtnat}
\bibliography{refs}

\end{document}